\documentclass[conference]{IEEEtran}
\usepackage{pdfpages}
\usepackage{amsmath}
\usepackage{amssymb}
\usepackage{graphicx}
\usepackage{caption}  

\begin{document}

\title{Cramér-Rao Bound for Direct Position Estimation in OFDM Based Cellular Systems}

\author{
    \IEEEauthorblockN{Sijia Li\textsuperscript{*}, Rui Sun\textsuperscript{*}, Bing Xu\textsuperscript{*}, and Yuanwei Liu\textsuperscript{\dag}}
    \IEEEauthorblockA{
        \textsuperscript{*}Department of Aeronautical and Aviation Engineering, The Hong Kong Polytechnic University, Hong Kong, China\\
        \textsuperscript{\dag}Department of Electrical and Electronic Engineering, The University of Hong Kong, Hong Kong, China\\
        E-mail: \{sijia-franz.li, rrrui.sun\}@connect.polyu.hk; pbing.xu@polyu.edu.hk; yuanwei@hku.hk}
}
\IEEEspecialpapernotice{(Invited Paper)}
\maketitle

\begin{abstract}
Although direct position estimation (DPE) has been demonstrated to offer enhanced robustness in GNSS receivers, its theoretical limits and performance in OFDM based positioning systems remain largely unexplored. In this paper, the Cramér-Rao bound (CRB) for DPE using OFDM based cellular signals is derived and benchmarked against the conventional two-step positioning method to assess their relative performance in non-line-of-sight (NLOS) dominated multipath environments. Numerical results reveal that 1) the DPE method consistently outperforms the two-step approach in OFDM systems under all evaluated conditions; 2) a large bandwidth is crucial in both methods, and increasing subcarrier spacing is more beneficial for a fixed bandwidth; 3) utilizing multiple OFDM symbols for positioning leads to substantial improvements in localization accuracy compared to relying on a single symbol. However, further increasing the number of symbols yields marginal improvements while significantly increasing computational complexity.
\end{abstract}

\IEEEpeerreviewmaketitle

\section{Introduction}
The global commercialization of the fifth-generation (5G) communication network has marked a milestone in wireless communications, while research and development into the sixth-generation (6G) technologies is already underway worldwide for facilitating new applications and services~\cite{ref1}. In particular, autonomous vehicles, including self-driving cars and unmanned aerial vehicles (UAVs), are expected to play an essential role in 6G-enabled intelligent transportation systems~\cite{ref2}. Precise localization is indispensable in maintaining operational safety, efficiency, and reliability for autonomous systems. This becomes especially crucial in densely populated urban areas, where the complexity of navigation and the existence of dynamic obstacles pose significant challenges.

As one of the most predominant localization technologies, the Global Navigation Satellite System (GNSS) faces substantial challenges in urbanized settings, where dense infrastructure and complex surroundings result in severe signal blockage, reflection, and multipath effects, thereby degrading its positioning accuracy~\cite{ref3}. This highlights the necessity of investigating complementary positioning techniques beyond GNSS to enable more reliable navigation for autonomous systems in urban canyons. The exploitation of orthogonal frequency-division multiplexing (OFDM) based cellular signals for positioning, navigation, and timing (PNT) has attracted considerable attention as a viable complement to GNSS in challenging environments~\cite{ref4},~\cite{ref5},~\cite{ref6},~\cite{ref7},~\cite{ref8}. On the one hand, the dense deployment of cellular infrastructure in urban environments ensures ubiquitous availability of cellular signals for navigation purposes. On the other hand, the large bandwidth of cellular signals allows for high ranging accuracy, thereby facilitating precise target localization. However, the major challenge in exploiting OFDM based cellular signals for positioning lies in their vulnerability to multipath and non-line-of-sight (NLOS) propagation channels, which are particularly common in urban street canyons. 

Direct position estimation (DPE) has been proposed as a more robust method in terms of achieving precise positioning in harsh propagation environments, which utilizes an idea of solving the receiver position, velocity, and timing (PVT) in a \emph{direct} manner via maximum likelihood (ML) in the navigation domain~\cite{ref9},~\cite{ref10}. Theoretical studies have demonstrated that DPE can achieve superior performance compared to traditional two-step positioning approaches under challenging conditions, such as multipath~\cite{ref11},~\cite{ref12}. However, most existing research on DPE has focused exclusively on GNSS signals, while its application to OFDM based cellular signals remains at a nascent stage. A recent study in~\cite{ref13} applies DPE to 5G positioning reference signals (PRS) and demonstrates improved performance over the conventional two-step time difference of arrival (TDOA) method in a NLOS dominated urbanized scenario. However, the fundamental positioning performance characterization of DPE in OFDM based cellular systems has not been explored yet. Driven by this motivation, this paper presents the derivation of the Cramér–Rao Bound (CRB) for the DPE method in OFDM systems and benchmarks its performance against the conventional two-step approach.

The rest of this paper is organized as follows. Section II presents the system model for the DPE positioning framework in OFDM systems. Section III derives the CRB for the DPE method based on OFDM signals. Section IV discusses the numerical results. Finally, Section V concludes the paper.
\section{System Model}
In this study, we consider a multi-source single-input-single-output (SISO) downlink OFDM system, where the positioning target receives the known downlink reference signals transmitted by multiple base stations (BSs) for localization. Without loss of generality, we assume that all BSs are perfectly synchronized while the user equipment (UE) has unknown clock bias that is to be estimated. In the following, we firstly present the OFDM based signal model in multipath environment. Then, we deliver the mathematical description of the corresponding DPE framework.
\subsection{Signal Model}
Consider an OFDM frame with $P$ symbols. Let $K$ denote the number of subcarriers, $T_{\text{s}}$ denote the elementary duration of an OFDM symbol, and $T_{\text{cp}}$ denote the length of the cyclic-prefix (CP). Therefore, the subcarrier-spacing, total signal bandwidth, and the overall duration of each OFDM symbol including CP are given by $\Delta f = 1/{T_\text{s}}$, $B = K\Delta f$, and $T_{\text{tot}} = T_{\text{s}} + T_{\text{cp}}$ respectively. Assuming that there are $M$ BSs contributing to the target localization, the baseband transmit signal from the $m$-th BS can be expressed as
\begin{equation}
    x^{m}(t) \!= \!\frac{1}{\sqrt{K}}\sum_{p=0}^{P-1} \sum_{k=0}^{K-1} d_{k,p}^{m} e^{j2\pi k \Delta f (t-pT_{\text{s}}-T_{\text{cp}})} \text{rect}\!\left(\frac{t-pT_{\text{s}}}{T_{\text{s}}}\right)\!,
\end{equation}
where $d_{k,p}^{m}$ denotes the complex symbol modulated onto the $k$-th subcarrier in the $p$-th OFDM symbol transmitted by the $m$-th BS, and $\text{rect}(t)$ represents the rectangular pulse function, which equals 1 if $t \in [0 , 1]$ and 0 otherwise.

For each OFDM symbol, the received signal is sampled at the sampling rate $D = N\Delta f$, where $N$ denotes the point of discrete Fourier transform (DFT). Without loss of generality, we assume $N \geq K$ to enable oversampling at the receiver side for a high multipath resolution. After discarding CP, the $n$-th $(n = 0,1,\ldots,N-1)$ sample in the $p$-th OFDM symbol can be expressed as
\begin{equation}
\begin{aligned}
    y_p[n] &= y_p(n/D) \\
    &= \!\frac{1}{\sqrt{K}}\!\!\sum_{m=1}^{M} \!\sum_{l=0}^{L^m\!\!-\!1}\! \sum_{k=0}^{K-1}
    h_{l,p}^{m} d_{k,p}^{m} e^{j2\pi kn/N} 
    e^{-j2\pi k \Delta f \tau_l^m(\boldsymbol{\gamma},\boldsymbol{\vartheta}^m)} \\
    &\quad\quad\quad\quad\quad\quad\quad\quad \times e^{j2\pi f_l^m(\boldsymbol{\gamma},\boldsymbol{\vartheta}^m)(n/D + T_{\text{cp}})} 
    + w_p[n],
\end{aligned}
\end{equation}
where $L^{m}$ is the number of multipath components comprised in the received signals that are either reflected or scattered in the environment, and $h_{l,p}^{m}$ is the complex channel coefficient of the $l$-th path in the $p$-th OFDM symbol received from the $m$-th BS. $w_p \sim \mathcal{CN}(0,\sigma^2)$ denotes the circularly symmetric complex additive white Gaussian noise (AWGN) with zero mean and variance $\sigma^2$. Multipath delays and Doppler shifts, denoted by $\tau_l^m(\boldsymbol{\gamma},\boldsymbol{\vartheta}^m)$ and $f_l^m(\boldsymbol{\gamma},\boldsymbol{\vartheta}^m)$, are modeled as $\tau_0^m(\boldsymbol{\gamma})$ and $f_0^m(\boldsymbol{\gamma})$ for line-of-sight (LOS) path ($l=0$), and as $\tau_0^m(\boldsymbol{\gamma}) + \Delta \tau_l^m$ and $f_l^m(\boldsymbol{\gamma}) + \Delta f_0^m$ for NLOS components ($l\geq1$). The vector $\boldsymbol{\vartheta}^m = \left[\Delta \tau_1^m,\ldots,\Delta\tau_{L^m-1}^m,\Delta f_1^m,\ldots,\Delta f_{L^m-1}^m\right]^T$ collects the NLOS biases in the received signal at the $m$-th BS, while $\boldsymbol{\gamma} = \left[\mathbf{p}^T , \mathbf{v}^T, \delta t\right]^T$ denotes the UE state parameter, which encapsulates the PVT information. Therefore, the time of arrival (TOA) and the Doppler shift of the LOS path can be respectively expressed as
\begin{align}
     \tau_{0}^{m}(\boldsymbol{\gamma}) &= \frac{1}{c} \Vert \mathbf{p}^{m} - \mathbf{p} \Vert + \delta t, \\
     f_{0}^{m}(\boldsymbol{\gamma}) &= \mathbf{v}^T \frac{\mathbf{p}^{m} - \mathbf{p}}{\Vert \mathbf{p}^{m} - \mathbf{p} \Vert} \frac{f_c}{c},
\end{align}
where $\mathbf{p} = [p_x, p_y, p_z]^T$ and $\mathbf{v} = [v_x, v_y, v_z]^T$ are the 3D position and velocity vectors respectively. $\delta t$ denotes the clock bias of the UE, $f_c$ denotes the carrier frequency, and $c$ is the speed of light.

Let $\mathbf{s}_{l,p}^m(\boldsymbol{\gamma},\boldsymbol{\vartheta}^m) \in \mathbb{C}^{N\times 1}$ denote the basis function corresponding to the $l$-th path received from the $m$-th BS, which can be expressed as
\begin{equation}
\begin{aligned}
    \mathbf{s}_{l,p}^m(\boldsymbol{\gamma},\boldsymbol{\vartheta}^m) &= \mathbf{x}_p^m\left(n/D-\tau_l^m(\boldsymbol{\gamma},\boldsymbol{\vartheta}^m)+T_{\text{cp}}\right)\\ &\quad \quad\quad\quad\quad\quad \quad\times e^{j2\pi f_l^m(\boldsymbol{\gamma},\boldsymbol{\vartheta}^m)(n/D+T_{\text{cp}})}.
\end{aligned}
\end{equation}
Therefore, the received signal for each OFDM symbol can be expressed in vector form as
\begin{equation}
    \mathbf{y}_p = \mathbf{S}_p(\boldsymbol{\gamma},\boldsymbol{\vartheta})\mathbf{h}_p + \mathbf{w}_p,
\end{equation}
where $\mathbf{h}_p \in \mathbb{C}^{Q \times 1}$ denotes the vector that collects all the complex channel coefficients for the $p$-th OFDM symbol, including both LOS and NLOS components. $Q = \sum_{m=1}^{M}L^{m}$ denotes the total number of signal replicas comprised in the received signal from all BS. $\mathbf{S}_p(\boldsymbol{\gamma},\boldsymbol{\vartheta}) \in \mathbb{C}^{N \times Q}$ denotes the basis matrix. Therefore, the overall received signal across the entire OFDM frame is expressed as $\mathbf{y} = \left[\mathbf{y}_0^T,\ldots,\mathbf{y}_{P-1}^T\right]^T$, which serves as the vector collecting all the received signals for the following estimation.

\subsection{DPE Framework in Downlink OFDM Systems}
We now derive the maximum likelihood estimation (MLE) for the UE state parameter based on the previously introduced signal model, where each received signal vector has a complex Gaussian distribution of $\mathbf{y}_p \sim \mathcal{CN}\left(\mathbf{S}_p(\boldsymbol{\gamma},\boldsymbol{\vartheta})\mathbf{h}_p, \sigma^2\mathbf{I}_N\right)$. Therefore, the log-likelihood function for estimating the unknown variables $\boldsymbol{\xi}=\big[\mathbf{h}^T,\boldsymbol{\vartheta}^T,\boldsymbol{\gamma}^T\big]^T$ from the entire vector $\mathbf{y}$ is given by
\begin{equation}
    \log f_{\mathbf{y}}\left(\mathbf{y};\boldsymbol{\xi}\right) \!=\! -PN\log\left(\pi \sigma^2\right) - \frac{1}{\sigma^2}\sum_{p=0}^{P-1}\Vert \mathbf{y}_p - \mathbf{S}_p(\boldsymbol{\gamma},\boldsymbol{\vartheta})\mathbf{h}_p \Vert^2.
\end{equation}
Accordingly, the ML estimates of the unknown variables that maximize the likelihood function are given by
\begin{equation}
    \hat{\boldsymbol{\xi}} = \arg\min_{\boldsymbol{\xi}}\sum_{p=0}^{P-1}\Vert \mathbf{y}_p-\mathbf{S}_p(\boldsymbol{\gamma},\boldsymbol{\vartheta})\mathbf{h}_p \Vert^2.
    \label{eq_MLE}
\end{equation}
For any given value of $\boldsymbol{\gamma}$ and $\boldsymbol{\vartheta}$, the optimal estimator of $\mathbf{h}_p$ in each OFDM symbol can be obtained as
\begin{equation}
    \begin{aligned}
        \hat{\mathbf{h}}_p &=\arg\min_{\mathbf{h}}\sum_{p=0}^{P-1}\Vert \mathbf{y}_p - \mathbf{S}_p(\boldsymbol{\gamma},\boldsymbol{\vartheta})\mathbf{h}_p\Vert^2\\
        &= \left(\mathbf{S}_p^{H}(\boldsymbol{\gamma},\boldsymbol{\vartheta})\mathbf{S}_p(\boldsymbol{\gamma},\boldsymbol{\vartheta})\right)^{-1}\mathbf{S}_p^{H}(\boldsymbol{\gamma},\boldsymbol{\vartheta})\mathbf{y}_p.
    \end{aligned}
\end{equation}
Substituting the above solution into~(\ref{eq_MLE}) yields
\begin{equation}
\begin{aligned}
    &\sum_{p=0}^{P-1}\Vert \mathbf{y}_p - \mathbf{S}_p(\boldsymbol{\gamma},\boldsymbol{\vartheta})\mathbf{h}_p \Vert^2 =\sum_{p=0}^{P-1}\Vert \mathbf{y}_p \Vert^2 \\&- \mathbf{y}_p^H\mathbf{S}_p(\boldsymbol{\gamma},\boldsymbol{\vartheta})\left(\mathbf{S}_p^H(\boldsymbol{\gamma},\boldsymbol{\vartheta})\mathbf{S}_p(\boldsymbol{\gamma},\boldsymbol{\vartheta})\right)^{-1}\mathbf{S}_p^H(\boldsymbol{\gamma},\boldsymbol{\vartheta})\mathbf{y}_p,
    \label{eq_costFun}
\end{aligned}
\end{equation}
which results in the equivalent estimator of $\boldsymbol{\gamma}$ given as the maximization of the non-negative cost function in \eqref{eq_costFun}.

A fine-resolution grid based search over the area of interest can be employed to solve the above maximization problem, and the computational complexity can be significantly offloaded through graphics processing unit (GPU) acceleration. This is because the cost function evaluated at different grid points are mutually independent and can be executed in parallel, which aligns well with the parallel processing capability of GPU~\cite{ref14}. 

\section{Cramér-Rao Bound Analysis}
In this section, we derive the CRB of the target localization in the DPE framework, which serves as the performance metric by providing a tight lower bound on the mean squared error (MSE) achievable by any unbiased estimator under general conditions~\cite{ref15}. As discussed in the previous section, it is clear that the received signal is complex white Gaussian, i.e., $\mathbf{y} \sim \mathcal{CN}\left(\mathbf{S}(\boldsymbol{\gamma},\boldsymbol{\vartheta})\mathbf{h},\sigma^2\mathbf{I}_{PN}\right)$. Let $\boldsymbol{\eta}=\big[\boldsymbol{\alpha}^T,\boldsymbol{\vartheta}^T, \boldsymbol{\gamma}^T\big]^T$ denote the real-valued parameters to estimate, where $\boldsymbol{\alpha} = \big[\Re\left\{\mathbf{h}\right\}^T,\Im\left\{\mathbf{h}\right\}^T\big]^T$ is the vector collecting the real part and the imaginary part of the complex channel coefficients. $\mathbf{u}(\boldsymbol{\eta})=\mathbf{S}(\boldsymbol{\gamma},\boldsymbol{\vartheta})\mathbf{h} \in \mathbb{C}^{PN\times1}$ denotes the observation vector. The element at the $i$-row and $j$-column of the Fisher information matrix (FIM) for estimating $\boldsymbol{\eta}$ is given by~\cite{ref16}
\begin{equation}
    \big[\mathbf{J}_{\boldsymbol{\eta}}\big]_{i,j} = 
    \frac{2}{\sigma^2}\Re\left\{
    \frac{\partial\mathbf{u}^H(\boldsymbol{\eta})}{\partial\eta_i}
    \frac{\partial\mathbf{u}(\boldsymbol{\eta})}{\partial\eta_j}
    \right\},
\end{equation}
where $\eta_i$ is the $i$-th element of $\boldsymbol{\eta}$, and the overall FIM can be expressed as
\begin{equation}
\mathbf{J}_{\boldsymbol{\eta}} = \frac{2}{\sigma^2}
\begin{bmatrix}
\mathbf{J}_{\boldsymbol{\alpha}\boldsymbol{\alpha}} & \mathbf{J}_{\boldsymbol{\alpha}\boldsymbol{\vartheta}} & \mathbf{J}_{\boldsymbol{\alpha}\boldsymbol{\gamma}}\\[0.5ex]
\mathbf{J}_{\boldsymbol{\alpha}\boldsymbol{\vartheta}}^T &
\mathbf{J}_{\boldsymbol{\vartheta}\boldsymbol{\vartheta}} &
\mathbf{J}_{\boldsymbol{\vartheta}\boldsymbol{\gamma}}\\[0.5ex]
\mathbf{J}_{\boldsymbol{\alpha}\boldsymbol{\gamma}}^T & 
\mathbf{J}_{\boldsymbol{\vartheta}\boldsymbol{\gamma}}^T &
\mathbf{J}_{\boldsymbol{\gamma}\boldsymbol{\gamma}}
\end{bmatrix}.
\label{eq_FIM}
\end{equation}
Let $\mathbf{\Xi}_{\boldsymbol{\alpha}} = \partial\mathbf{u}(\boldsymbol{\eta})/\partial\boldsymbol{\alpha} = \big[\mathbf{S}(\boldsymbol{\gamma},\boldsymbol{\vartheta}), j\mathbf{S}(\boldsymbol{\gamma},\boldsymbol{\vartheta})\big]$, $\mathbf{\Xi}_{\boldsymbol{\vartheta}} = \partial\mathbf{u}(\boldsymbol{\eta})/\partial\boldsymbol{\vartheta}$, and $\mathbf{\Xi}_{\boldsymbol{\gamma}} = \partial\mathbf{u}(\boldsymbol{\eta})/\partial\boldsymbol{\gamma}$. Therefore, the elements of each submatrix in~(\ref{eq_FIM}) can be obtained as
\begin{equation}
    \mathbf{J}_{\boldsymbol{\alpha}\boldsymbol{\alpha}} = 
    \begin{bmatrix}
        \Re\left\{\mathbf{S}^H(\boldsymbol{\gamma},\boldsymbol{\vartheta})\mathbf{S}(\boldsymbol{\gamma},\boldsymbol{\vartheta})\right\} & 
        -\Im\left\{\mathbf{S}^H(\boldsymbol{\gamma},\boldsymbol{\vartheta})\mathbf{S}(\boldsymbol{\gamma},\boldsymbol{\vartheta})\right\} \\[0.5ex]
        \Im\left\{\mathbf{S}^H(\boldsymbol{\gamma},\boldsymbol{\vartheta})\mathbf{S}(\boldsymbol{\gamma},\boldsymbol{\vartheta})\right\} &
        \Re\left\{\mathbf{S}^H(\boldsymbol{\gamma},\boldsymbol{\vartheta})\mathbf{S}(\boldsymbol{\gamma},\boldsymbol{\vartheta})\right\}
    \end{bmatrix},
    \label{eq_Jaa}
\end{equation}
\begin{equation}
    \mathbf{J}_{\boldsymbol{\alpha}\boldsymbol{\vartheta}} =
    \begin{bmatrix}
        \Re\left\{\mathbf{S}^H(\boldsymbol{\gamma},\boldsymbol{\vartheta})\mathbf{\Xi}_{\boldsymbol{\vartheta}}\right\} \\[0.5ex]
        \Im\left\{\mathbf{S}^H(\boldsymbol{\gamma},\boldsymbol{\vartheta})\mathbf{\Xi}_{\boldsymbol{\vartheta}}\right\}
    \end{bmatrix},
    \mathbf{J}_{\boldsymbol{\alpha}\boldsymbol{\gamma}} =
    \begin{bmatrix}
        \Re\left\{\mathbf{S}^H(\boldsymbol{\gamma},\boldsymbol{\vartheta})\mathbf{\Xi}_{\boldsymbol{\gamma}}\right\} \\[0.5ex]
        \Im\left\{\mathbf{S}^H(\boldsymbol{\gamma},\boldsymbol{\vartheta})\mathbf{\Xi}_{\boldsymbol{\gamma}}\right\}
    \end{bmatrix},
    \label{eq_Jatag}
\end{equation}
\begin{equation}
    \mathbf{J}_{\boldsymbol{\vartheta}\boldsymbol{\vartheta}} \!=\!
    \Re\left\{
    \mathbf{\Xi}_{\boldsymbol{\vartheta}}^H\mathbf{\Xi}_{\boldsymbol{\vartheta}}
    \right\}\!,
    \mathbf{J}_{\boldsymbol{\vartheta}\boldsymbol{\gamma}} \!=\!
    \Re\left\{
    \mathbf{\Xi}_{\boldsymbol{\vartheta}}^H\mathbf{\Xi}_{\boldsymbol{\gamma}}
    \right\}\!,
    \mathbf{J}_{\boldsymbol{\gamma}\boldsymbol{\gamma}} \!=\!
    \Re\left\{
    \mathbf{\Xi}_{\boldsymbol{\gamma}}^H\mathbf{\Xi}_{\boldsymbol{\gamma}}
    \right\}.
    \label{eq_Jtttggg}
\end{equation}
Prior to expanding the expression of the above submatrices, we first define the transform matrix as
\begin{equation}
\begin{aligned}
      \mathbf{\Theta}(\boldsymbol{\gamma},\boldsymbol{\vartheta}) &\in \mathbb{C}^{N\times QK}\\
      &\!\!\!\!\!\!\!\!\!\!\!\!\!= \left[\mathbf{\Upsilon}_1(\boldsymbol{\gamma},\boldsymbol{\vartheta})\mathbf{F}_N^H\mathbf{\Gamma}_1(\boldsymbol{\gamma},\boldsymbol{\vartheta}),\ldots,\mathbf{\Upsilon}_Q(\boldsymbol{\gamma},\boldsymbol{\vartheta})\mathbf{F}_N^H\mathbf{\Gamma}_Q(\boldsymbol{\gamma},\boldsymbol{\vartheta}))
    \right],
\end{aligned}
\end{equation}
where
$\mathbf{F}_N \in \mathbb{C}^{K\times N}$ is the oversampled N-point DFT matrix, 
with each element given by $\left[\mathbf{F}_N\right]_{u,v} = (1/\sqrt{K}) e^{-j2\pi uv/N}$. 
The Doppler shift matrix is given by 
$\mathbf{\Upsilon}_{q}(\boldsymbol{\gamma},\boldsymbol{\vartheta}) = 
\mathrm{diag}\!\left(e^{j2\pi f_q(\boldsymbol{\gamma},\boldsymbol{\vartheta})T_{\mathrm{cp}}},\ldots,
e^{j2\pi f_q(\boldsymbol{\gamma},\boldsymbol{\vartheta})((N\!{-}\!1)/D{+}T_{\mathrm{cp}})}\right) 
\in \mathbb{C}^{N\times N}$.
The multipath delay matrix is denoted by
$\mathbf{\Gamma}_{q}(\boldsymbol{\gamma},\boldsymbol{\vartheta}) = 
\mathrm{diag}\!\left(1,\ldots,
e^{-j2\pi (K{-}1)\Delta f\, \tau_q(\boldsymbol{\gamma},\boldsymbol{\vartheta})}\right) 
\in \mathbb{C}^{K\times K}$.
Then, define the matrix of partial derivatives with respect to the delay and Doppler shift parameters as
\begin{equation}
     \begin{aligned}
         \mathbf{\Psi}(\boldsymbol{\gamma},\boldsymbol{\vartheta}) &=\left[\mathbf{\Psi}_{\boldsymbol{\gamma}}(\boldsymbol{\gamma},\boldsymbol{\vartheta}),\mathbf{\Psi}_{\boldsymbol{\vartheta}}(\boldsymbol{\gamma},\boldsymbol{\vartheta})\right] \in \mathbb{C}^{N\times2QK}\\
         &\!\!\!\!\!\!\!\!\!\!\!\!\!\!\!\!=\left[
         \partial_f\!\mathbf{\Upsilon}_1(\boldsymbol{\gamma},\boldsymbol{\vartheta})\mathbf{F}_N^H\mathbf{\Gamma}_1(\boldsymbol{\gamma},\boldsymbol{\vartheta}),\mathbf{\Upsilon}_1(\boldsymbol{\gamma},\boldsymbol{\vartheta})\mathbf{F}_N^H\partial_\tau\mathbf{\Gamma}_1(\boldsymbol{\gamma},\boldsymbol{\vartheta}),\ldots,\right.\\
         &\quad\!\!\!\!\!\!\!\!\!\!\!\!\!\left.\partial_f\!\mathbf{\Upsilon}_M(\boldsymbol{\gamma},\boldsymbol{\vartheta})\mathbf{F}_N^H\mathbf{\Gamma}_M(\boldsymbol{\gamma},\boldsymbol{\vartheta}),\mathbf{\Upsilon}_M(\boldsymbol{\gamma},\boldsymbol{\vartheta})\mathbf{F}_N^H\partial_\tau\mathbf{\Gamma}_M(\boldsymbol{\gamma},\boldsymbol{\vartheta}),\right.\\
         &\!\!\!\!\!\!\!\!\!\!\!\!\!\!\!\!\!\!\!\!\!\!\left.\partial_f\!\mathbf{\Upsilon}\!_{M\!+\!1}(\boldsymbol{\gamma},\boldsymbol{\vartheta})\mathbf{F}_N^H\mathbf{\Gamma}_{M\!+\!1}(\boldsymbol{\gamma},\boldsymbol{\vartheta}),\mathbf{\Upsilon}\!_{M\!+\!1}(\boldsymbol{\gamma},\boldsymbol{\vartheta})\mathbf{F}_N^H\partial_\tau\mathbf{\Gamma}_{M\!+\!1}(\boldsymbol{\gamma},\boldsymbol{\vartheta}),\right.\\
         &\!\!\!\!\!\!\!\!\!\!\!\!\!\!\!\!\left.\ldots,\partial_f\!\mathbf{\Upsilon}_Q(\boldsymbol{\gamma},\boldsymbol{\vartheta})\mathbf{F}_N^H\mathbf{\Gamma}_Q(\boldsymbol{\gamma},\boldsymbol{\vartheta}),\mathbf{\Upsilon}_Q(\boldsymbol{\gamma},\boldsymbol{\vartheta})\mathbf{F}_N^H\partial_\tau\mathbf{\Gamma}_Q(\boldsymbol{\gamma},\boldsymbol{\vartheta})
         \right],
     \end{aligned}
\end{equation}
where $\partial_f\!\mathbf{\Upsilon}_q(\boldsymbol{\gamma},\boldsymbol{\vartheta})$ and $\partial_\tau\mathbf{\Gamma}_q(\boldsymbol{\gamma},\boldsymbol{\vartheta})$ can be obtained as
\begin{equation}
\begin{aligned}
    \partial_f\!\mathbf{\Upsilon}_q(\boldsymbol{\gamma},\boldsymbol{\vartheta}) = 
    \operatorname{diag}\Big(j2\pi T_{\text{cp}}e^{j2\pi f_q(\boldsymbol{\gamma},\boldsymbol{\vartheta})T_{\text{cp}}}, \ldots,\quad\quad\quad\quad\quad\quad\\ 
    j2\pi ((N-1)/D+T_{\text{cp}})e^{j2\pi f_q(\boldsymbol{\gamma},\boldsymbol{\vartheta})((N-1)/D+T_{\text{cp}})}\Big),
\end{aligned}
\end{equation}
\begin{equation}
\begin{aligned}
    \partial_\tau\mathbf{\Gamma}_q(\boldsymbol{\gamma},\boldsymbol{\vartheta}) = 
    \operatorname{diag}\Big(0, \ldots,
    {-}j2\pi(K{-}1)\Delta fe^{{-}j2\pi \Delta f\tau_q(\boldsymbol{\gamma},\boldsymbol{\vartheta})}\Big).
\end{aligned}
\end{equation}
The matrix of partial derivatives with respect to the UE state parameter is given by
\begin{equation}
\mathbf{\Phi}(\boldsymbol{\gamma},\boldsymbol{\vartheta}) = 
\begin{bmatrix}
\left[\partial_{\boldsymbol{\gamma}}f_1(\boldsymbol{\gamma},\boldsymbol{\vartheta})^T,\partial_{\boldsymbol{\gamma}}\tau_1(\boldsymbol{\gamma},\boldsymbol{\vartheta})^T\right]^T \\
\vdots \\
\left[\partial_{\boldsymbol{\gamma}}f_M(\boldsymbol{\gamma},\boldsymbol{\vartheta})^T,\partial_{\boldsymbol{\gamma}}\tau_M(\boldsymbol{\gamma},\boldsymbol{\vartheta})^T\right]^T
\end{bmatrix}
\in \mathbb{C}^{2M\times7},
\end{equation}
where                                                                            
\begin{equation}
    \partial_{\gamma}f_q(\boldsymbol{\gamma},\boldsymbol{\vartheta})= \left[(\mathbf{v}_q-\mathbf{v})^T\frac{\mathbf{I}_3-\mathbf{g}_q\mathbf{g}_q^T}{\Vert\mathbf{p}_q-\mathbf{p}\Vert}\frac{f_c}{c},\mathbf{g}_q^T\frac{f_c}{c},0\right],
\end{equation}
\begin{equation}
    \partial_\gamma\tau_q(\boldsymbol{\gamma},\boldsymbol{\vartheta})=
    \left[-\frac{\mathbf{g}_q^T}{c},0,0,0,1\right],
\end{equation}
where $\mathbf{g}_q = \frac{\mathbf{p}_q-\mathbf{p}}{\Vert\mathbf{p}_q-\mathbf{p}\Vert}$ denotes the unitary direction vector of the $q$-th BS relative to the UE. Then, define the positioning signal matrix as
\begin{equation}
    \mathbf{D} = \operatorname{blkdiag}\left(
    \mathbf{I}_{L^1}\otimes \mathbf{d}_1,\ldots,\mathbf{I}_{L^M}\otimes\mathbf{d}_M
    \right),
\end{equation}
\begin{equation}
        \mathbf{D}_{\boldsymbol{\gamma}} = \operatorname{blkdiag}\left(
        \mathbf{I}_2\otimes\mathbf{d}_1,\ldots,\mathbf{I}_2\otimes\mathbf{d}_M
        \right),
\end{equation}
\begin{equation}
     \mathbf{D}_{\boldsymbol{\vartheta}} = \operatorname{blkdiag}\left(
    \mathbf{I}_{L^1\!{-}\!1}\otimes\left(\mathbf{I}_2\otimes\mathbf{d}_1\right),\ldots,\mathbf{I}_{L^M\!{-}\!1}\otimes\left(\mathbf{I}_2\otimes\mathbf{d}_M\right)
    \right),
\end{equation}
where $\otimes$ is the Kronecker product, $\mathbf{d}_q \in \mathbb{C}^{K\times 1}$ denotes the vector that collects the modulated data across all subcarriers and OFDM symbols. Finally, the channel coefficient matrices are defined as
\begin{figure*}[!t]
  \centering

  \begin{minipage}[t]{0.32\textwidth} 
    \centering
    \includegraphics[width=\linewidth]{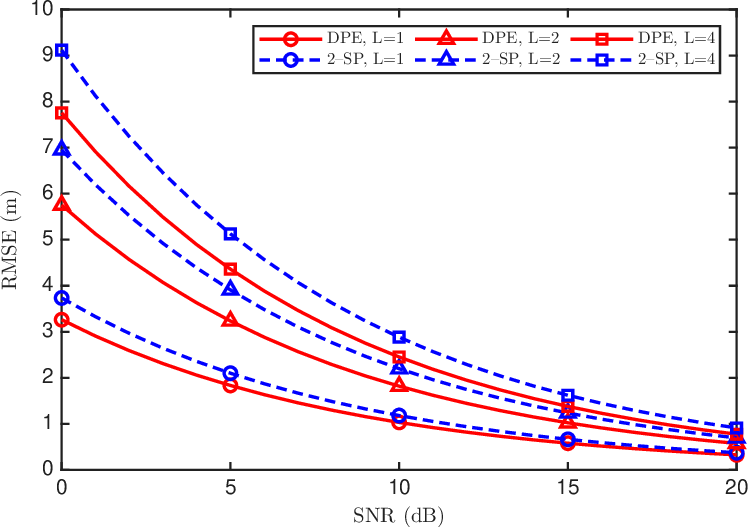}
    \captionsetup{type=figure}
    \caption{Comparison between the CRBs of the DPE method and the conventional two-step approach.}
    \label{fig:crb_snr}
  \end{minipage}\hfill
  \begin{minipage}[t]{0.32\textwidth}  
    \centering
    \includegraphics[width=\linewidth]{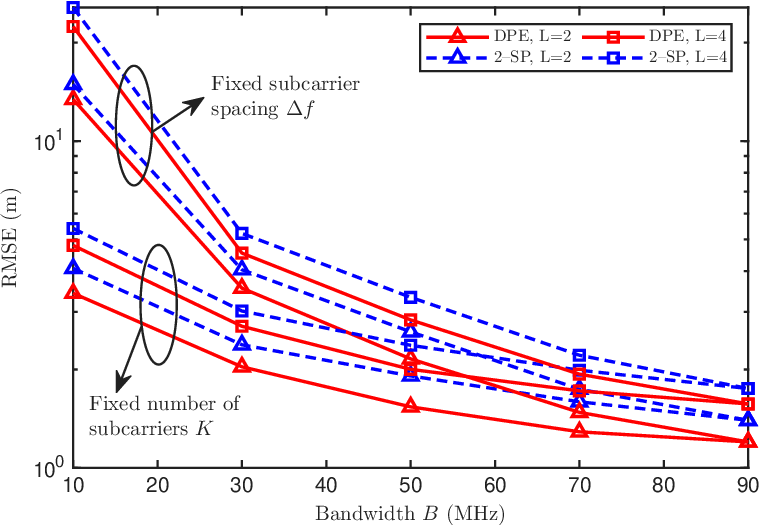}
    \captionsetup{type=figure}
    \caption{CRBs versus the bandwidth when fixing subcarrier spacing and the number of subcarriers.}
    \label{fig:crb_bw}
  \end{minipage}\hfill
  \begin{minipage}[t]{0.32\textwidth}  
    \centering
    \includegraphics[width=\linewidth]{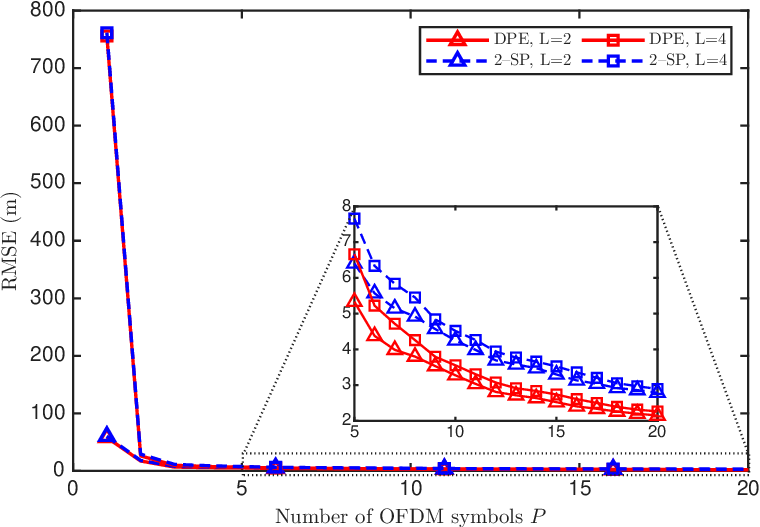}
    \captionsetup{type=figure}
    \caption{CRBs versus the number of OFDM symbols contributing to the positioning.}
    \label{fig:crb_nsym}
  \end{minipage}

\end{figure*}

\begin{equation}
    \mathbf{H}_{\boldsymbol{\gamma}} = 
    \begin{bmatrix}
        \operatorname{blkdiag}\left(\mathbf{I}_2\otimes h_{1,1},\ldots,\mathbf{I}_2\otimes h_{M,1}\right)\\
        \vdots\\
        \operatorname{blkdiag}\left(\mathbf{I}_2\otimes h_{1,P},\ldots,\mathbf{I}_2\otimes h_{M,P}\right)
    \end{bmatrix},
\end{equation}
\begin{equation}
\mathbf{H}_{\boldsymbol{\vartheta}} =
    \begin{bmatrix}
        \operatorname{blkdiag}\left(\mathbf{I}_2\otimes h_{M{+}1,1},\ldots,\mathbf{I}_2\otimes h_{Q,1}\right)\\
        \vdots\\
        \operatorname{blkdiag}\left(\mathbf{I}_2\otimes h_{M{+}1,P},\ldots,\mathbf{I}_2\otimes h_{Q,P}\right)
    \end{bmatrix}.
\end{equation}
Assuming identical basis matrices across all OFDM symbols in a frame, we obtain
\begin{equation}
    \mathbf{S}(\boldsymbol{\gamma},\boldsymbol{\vartheta}) = \mathbf{I}_p\otimes\mathbf{\Theta}(\boldsymbol{\gamma},\boldsymbol{\vartheta})\mathbf{D},
    \label{eq_28}
\end{equation}
\begin{equation}
    \mathbf{\Xi}_{\boldsymbol{\vartheta}} = \left(\mathbf{I}_p\otimes\mathbf{\Psi}_{\boldsymbol{\gamma}}(\boldsymbol{\gamma},\boldsymbol{\vartheta})\mathbf{D}_{\boldsymbol{\gamma}}\right)\mathbf{H}_{\boldsymbol{\vartheta}},
\end{equation}
\begin{equation}
    \mathbf{\Xi}_{\boldsymbol{\gamma}} = \left(\mathbf{I}_p\otimes\mathbf{\Psi}_{\boldsymbol{\gamma}}(\boldsymbol{\gamma},\boldsymbol{\vartheta})\mathbf{D}_{\boldsymbol{\gamma}}\right)\mathbf{H}_{\boldsymbol{\gamma}}\mathbf{\Phi}(\boldsymbol{\gamma},\boldsymbol{\vartheta}).
    \label{eq_30}
\end{equation}
By substituting \eqref{eq_28}--\eqref{eq_30} into \eqref{eq_Jaa}--\eqref{eq_Jtttggg}, the CRB for estimating the UE state parameter $\boldsymbol{\gamma}$ can be calculated as
\begin{equation}
    \text{CRB}(\boldsymbol{\gamma})\! =\! \left(\!\mathbf{J}_{\boldsymbol{\gamma}\boldsymbol{\gamma}} \!-\! \begin{bmatrix}
        \mathbf{J}_{\boldsymbol{\alpha}\boldsymbol{\gamma}}\\\mathbf{J}_{\boldsymbol{\vartheta}\boldsymbol{\gamma}}
    \end{bmatrix}^T\!\!\begin{bmatrix}
        \mathbf{J}_{\boldsymbol{\alpha}\boldsymbol{\alpha}} & \mathbf{J}_{\boldsymbol{\alpha}\boldsymbol{\vartheta}}\\
        \mathbf{J}^T_{\boldsymbol{\alpha}\boldsymbol{\vartheta}} & \mathbf{J}_{\boldsymbol{\vartheta}\boldsymbol{\vartheta}}
    \end{bmatrix}^{-1}\!\!\begin{bmatrix}
        \mathbf{J}_{\boldsymbol{\alpha}\boldsymbol{\gamma}}\\\mathbf{J}_{\boldsymbol{\vartheta}\boldsymbol{\gamma}}
    \end{bmatrix}\right)^{-1}\!\!\!\in \mathbb{R}^{7\times7}.
\end{equation}

\section{Numerical results}
In this section, numerical results are provided to demonstrate the derived CRB results. Unless otherwise specified, we set $f_c= 3.5$ GHz, $B = 5$ MHz, $\Delta f = 30$ KHz, $M = 4$, $P = 4$, $K = 128$, and $N = 2048$ for the simulation. In addition, we set the same number of multipaths for all the BSs, i. e., $L^1 = L^2 = \ldots = L^M = L$. The signal-to-noise ratio (SNR) is set to be 0 dB, and the signal-to-multipath ratio (SMR) is set to be -3 dB to model an NLOS propagation scenario. The target is considered to be non-static with a velocity of $\mathbf{v} = [3,5,0]^T$ and positioned at the origin. The positioning performance is evaluated in terms of the 3D root-MSE (RMSE). The CRB for the DPE method is compared to the CRB for the conventional two-step approach, which utilizes the ranging information of the received signal collected from each BS. The CRB of the target position using the conventional two-step approach is given by \cite{ref11}
\begin{equation}
\begin{aligned}
    \text{CRB}(\mathbf{p},\delta t) &= c^2\left(\left(\mathbf{T}^H\mathbf{W}\mathbf{T}\right)^{-1}\mathbf{T}^H\mathbf{W}\right)\mathbf{J}_{\tau\tau}^{-1}\\
    &\quad\quad\quad\quad\quad\left(\left(\mathbf{T}^H\mathbf{W}\mathbf{T}\right)^{-1}\mathbf{T}^H\mathbf{W}\right)^T \in \mathbb{R}^{4\times4},
\end{aligned}
\end{equation}
where $\mathbf{W} = \mathbf{I}$ is the weighting matrix, and 
\begin{equation}
    \mathbf{T} = \begin{bmatrix}
        \frac{x_1-\hat{x}}{\Vert\mathbf{p}_1-\hat{\mathbf{p}}\Vert} &
        \frac{y_1-\hat{y}}{\Vert\mathbf{p}_1-\hat{\mathbf{p}}\Vert} &
        \frac{z_1-\hat{z}}{\Vert\mathbf{p}_1-\hat{\mathbf{p}}\Vert} &
        1\\
        \vdots & \vdots & \vdots & \vdots\\
        \frac{x_M-\hat{x}}{\Vert\mathbf{p}_M-\hat{\mathbf{p}}\Vert} &
        \frac{y_M-\hat{x}}{\Vert\mathbf{p}_M-\hat{\mathbf{p}}\Vert} &
        \frac{z_M-\hat{x}}{\Vert\mathbf{p}_M-\hat{\mathbf{p}}\Vert} &
        1
    \end{bmatrix}.
\end{equation}
$\mathbf{J}_{\tau\tau}$ is the submatrix of $\mathbf{J}_{\boldsymbol{v}\boldsymbol{v}}$, where $\boldsymbol{v} = \big[\boldsymbol{\tau}_{\text{LOS}}^T,\mathbf{f}^T_{\text{LOS}}\big]^T$ denotes the vector collecting the TOAs and the Doppler shifts of the LOS components. The derivation of $\mathbf{J}_{\tau\tau}$ in multipath propagation channel is well studied in \cite{ref17}.

Fig.~\ref{fig:crb_snr} evaluates the positioning performance of the DPE method and the conventional two-step approach in multipath propagation scenarios. It can be observed that for multipath scenarios ($L=2$ and $L=4$), the RMSEs become significantly higher than the LOS condition. As expected, the DPE method demonstrates superior performance over the two-step approach in OFDM systems, especially in scenarios with low SNR and multipath propagation. 

Fig.~\ref{fig:crb_bw} studies the impact of the bandwidth under the conditions of fixed subcarrier spacing and a fixed number of subcarriers. It aligns with the expectation that the bandwidth has a considerable effect on positioning performance. Interestingly, a greater enhancement in positioning accuracy can be achieved by increasing the subcarrier spacing while maintaining a constant number of subcarriers, highlighting the importance of subcarrier separation in addition to total bandwidth.

Fig.~\ref{fig:crb_nsym} explores the impact of the number of OFDM symbols on the positioning performance. It is observed that when only a limited number of symbols is available, both approaches fail to provide reliable positioning when multipath is severe, whereas having more OFDM symbols for positioning leads to better accuracy. However, increasing the number of OFDM symbols beyond a certain point yields only marginal performance gains while significantly increasing computational complexity.
\section{Conclusion}
This study investigates the performance of the DPE method in an OFDM based cellular system. The CRB for target localization within the OFDM-DPE framework is derived and analyzed with respect to key system parameters. Numerical results show that the DPE method consistently outperforms the conventional two-step approach under all evaluated conditions. In particular, bandwidth plays a critical role in positioning accuracy, and increasing subcarrier spacing leads to better performance. Additionally, utilizing multiple OFDM symbols enhances positioning accuracy, whereas excessive increases provide marginal improvements
at the cost of higher computational complexity.

\end{document}